\documentclass[a4paper,11pt]{article}
\usepackage{pos,amssymb}
\usepackage{comment}

\title{Commissioning of the camera of the first Large Size Telescope of the Cherenkov Telescope Array }
 \ShortTitle{Commissioning of the camera of LST1}
\author*[a]{T.~Saito}
\author[b]{C.~Delgado}
\author[c]{O.~Blanch}
\author[c]{M.~Artero}
\author[d]{J.~A. Barrio}
\author[e]{F.~Cassol}
\author[b]{C.~Diaz}
\author[a]{D.~Hadasch}
\author[e]{D.~Hoffmann}
\author[e]{J.~Houles}
\author[a]{Y.~Inome}
\author[f]{M.~Iori}
\author[c]{L.~Jouvin}
\author[c]{D.~Kerszberg}
\author[a]{Y.~Kobayashi}
\author[g]{H.~Kubo}
\author[b]{G.~Martinez}
\author[a]{D.~Mazin}
\author[c]{E.~Moretti}
\author[h]{T.~Nakamori}
\author[g]{S.~Nozaki}
\author[g]{T.~Oka}
\author[i]{A.~Okumura}
\author[f]{M.~Palatiello}
\author[b]{M.~Polo}
\author[k]{J.~Prast}
\author[a]{S.~Sakurai}
\author[j]{Y.~Sunada}
\author[a]{M.~Takahashi}
\author[a]{R.~Takeishi}
\author[d]{L.~A. Tejedor}
\author[l]{T.~Yamamoto}
\author[m]{T.~Yoshida}
\affiliation[a]{ICRR, University of Tokyo, 5-1-5, Kashiwa-no-ha, Kashiwa, Chiba 277-8582, Japan}
\affiliation[b]{CIEMAT, Avda. Complutense 40, 28040 Madrid, Spain}
\affiliation[c]{Institut de Fisica d'Altes Energies (IFAE), The Barcelona Institute of Science and Technology, Campus UAB, 08193 Bellaterra (Barcelona), Spain}
\affiliation[d]{EMFTEL department and IPARCOS, Universidad Complutense de Madrid, 28040 Madrid, Spain}
\affiliation[e]{Aix-Marseille Université, CNRS/IN2P3, CPPM, 163 Avenue de Luminy, 13288 Marseille cedex 09, France}
\affiliation[f]{INFN Sezione di Roma La Sapienza, P.le Aldo Moro, 2 - 00185 Roma, Italy}
\affiliation[g]{Graduate School of Science, Kyoto University, Sakyo-ku, Kyoto, 606-8502, Japan}
\affiliation[h]{Department of Physics, Yamagata University, Yamagata, Yamagata 990-8560, Japan}
\affiliation[i]{ISEE, Nagoya University, Chikusa-ku, Nagoya 464-8602, Japan}
\affiliation[j]{Grad. Sch. of Sci. and Eng., Saitama University, 255 Simo-Ohkubo, Sakura-ku, Saitama 338-8570, Japan}
\affiliation[k]{CNRS, Laboratoire d'Annecy de Physique des Particules - IN2P3, 74000 Annecy, France}
\affiliation[l]{Department of Physics, Konan University, Kobe, Hyogo, 658-8501, Japan}
\affiliation[m]{Faculty of Science, Ibaraki University, Mito, Ibaraki, 310-8512, Japan}
\forColl{CTA LST} 

\emailAdd{tsaito@icrr.u-tokyo.ac.jp}
\emailAdd{carlos.delgado@ciemat.es}
\emailAdd{blanch@ifae.es}

\abstract{The first Large Size Telescope (LST-1) of the Cherenkov Telescope Array has been operational since October 2018 at La Palma, Spain. 
   We report on the results obtained during the camera commissioning. The noise level of the readout is determined as a 0.2 p.e. level. The gain of PMTs are well equalized within 2\% variation, using the calibration flash system. The effect of the night sky background on the signal readout noise as well as the PMT gain estimation are also well evaluated. Trigger thresholds are optimized for the lowest possible gamma-ray energy threshold and the trigger distribution synchronization has been achieved within 1~ns precision. Automatic rate control realizes the stable observation with 1.5\% rate variation over 3 hours. The performance of the novel DAQ system demonstrates a less than 10\% dead time for 15 kHz trigger rate even with sophisticated online data correction.
}

\FullConference{37$^{\rm{th}}$ International Cosmic Ray Conference (ICRC 2021)\\
		July 12th -- 23rd, 2021\\
		Online -- Berlin, Germany}

\def\Gbps{\ifmmode{\rm Gbps}\else Gbps\fi}

\begin{document}
\maketitle

\section{Overview of the LST Camera}

\begin{figure}
    \centering
    \includegraphics[width=0.8\textwidth]{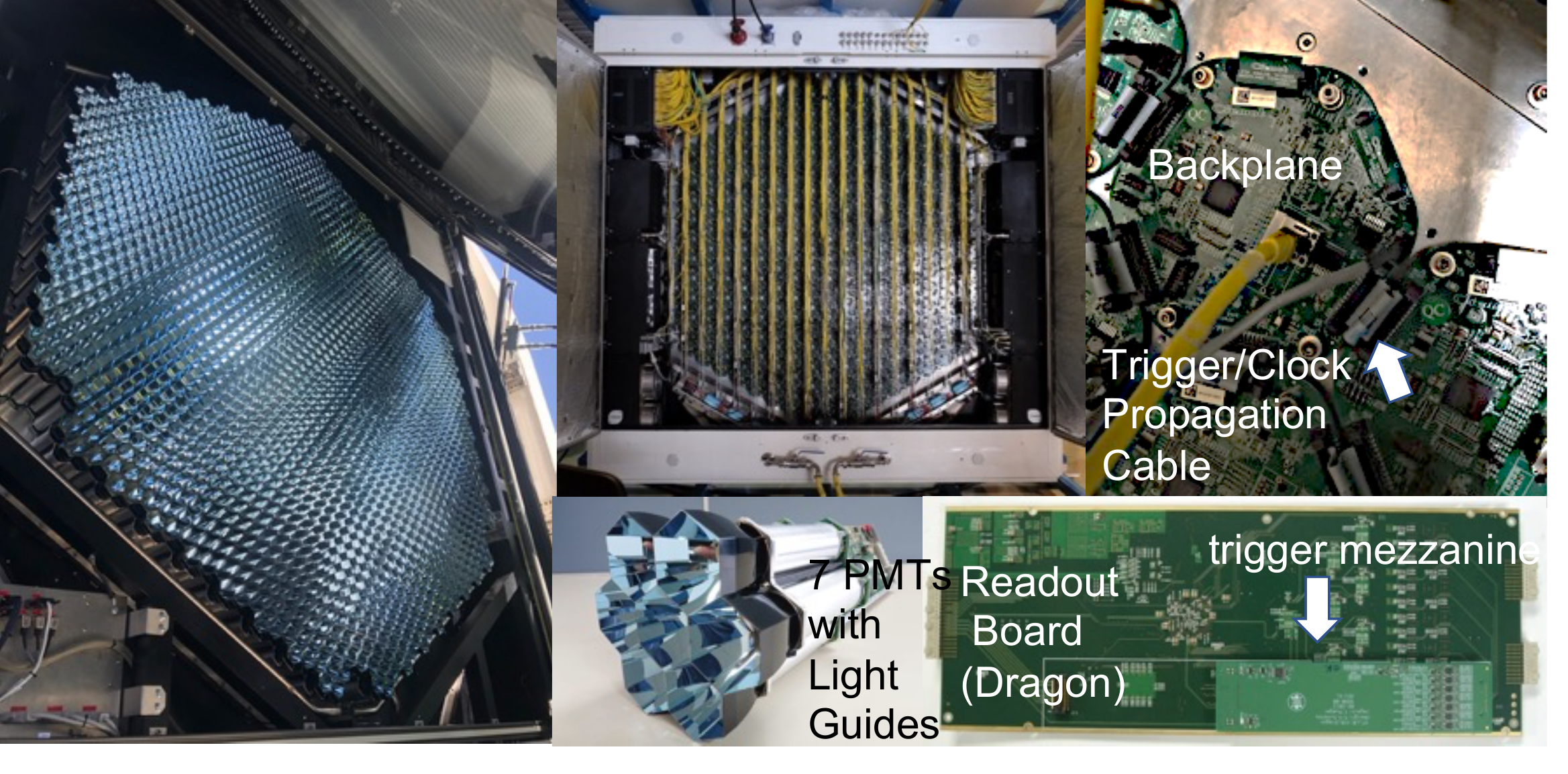}
    \caption{LST Camera (Left). It consists of 265 modules, each of which is equipped with 7 PMTs (bottom center). An analog trigger system (bottom right) is implemented where PMT signals from 3 modules (21 pixels) are summed up before a discriminator. The readout trigger signal is distributed through backplanes (top right).}
    \label{fig:camera}
\end{figure}
 
  The camera of LST consists of 265 modules, each of which is equipped with 7 PMTs (Fig.~\ref{fig:camera}). The PMTs have a quantum efficiency of about 40\% at 350 nm and the pulse width is about 3 ns. The output signal is amplified with 2 different gains so that a wide dymamic range (1 to 3000 photoelectrons) can be covered by the readout electronics. An analog trigger system is implemented where PMT signals from 3 modules (21 pixels) are summed up before a discriminator. If the summed signal exceeds the threshold, the digital trigger signal is generated and propagated through backplanes (BPs) to the trigger interface board (TIB) installed at the top left corner of the camera. TIB examines the state of readout system and the coincidence with neighboring telescopes, and if conditions are met, the readout trigger is sent to each modules through BPs.
Upon the triggers, the events are readout with a high speed data acquisition (DAQ) system with a 60 Gbps bandwidth. 

The camera body of LST has a square shape with the edge of about 3~m. Photo-sensitive area has a hexagonal shape and the field of view is about 4.5 degree in diameter. Each of 1855 pixels covers roughly 0.1 degrees. The camera is also equipped with a movable star imaging screen, where star images can be focused on, to calibrate mirror alignment and telescope pointing. 
Its movement was well studied and tested during the commissioning. 
 During operation, total power consumption amounts to 11 kW. Cooling liquid is circulated inside the module holder to cool down each modules. In addition, air circulation fans are equipped at the top of the camera which equalize the temperature over the camera. Cooling and electrical connectivity were well tested since the beginning of the commissioning.


\section{Noise Level}
We evaluated the noise level under various conditions and its stability. For this evaluation, dedicated pedestal data were taken using a pseudo-random trigger generated on the readout board. Pedestal charges are calculated as the charge extracted from a fixed 8~ns readout window.
Hereafter, the noise level is defined as the standard deviation of the pedestal charges of each pixel.

\begin{figure}
    \centering
    \includegraphics[width=0.35\textwidth]{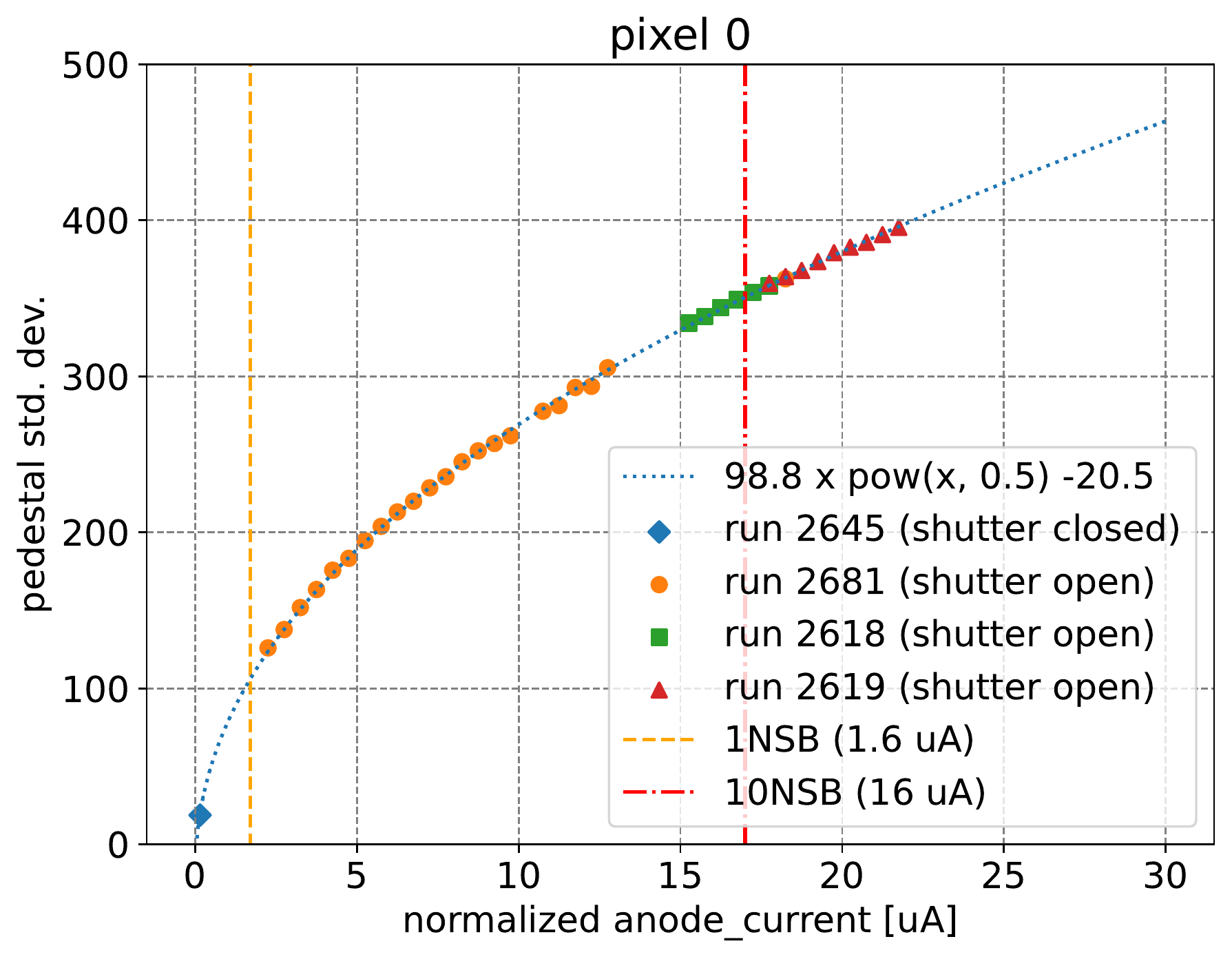}
    \includegraphics[width=0.55\textwidth]{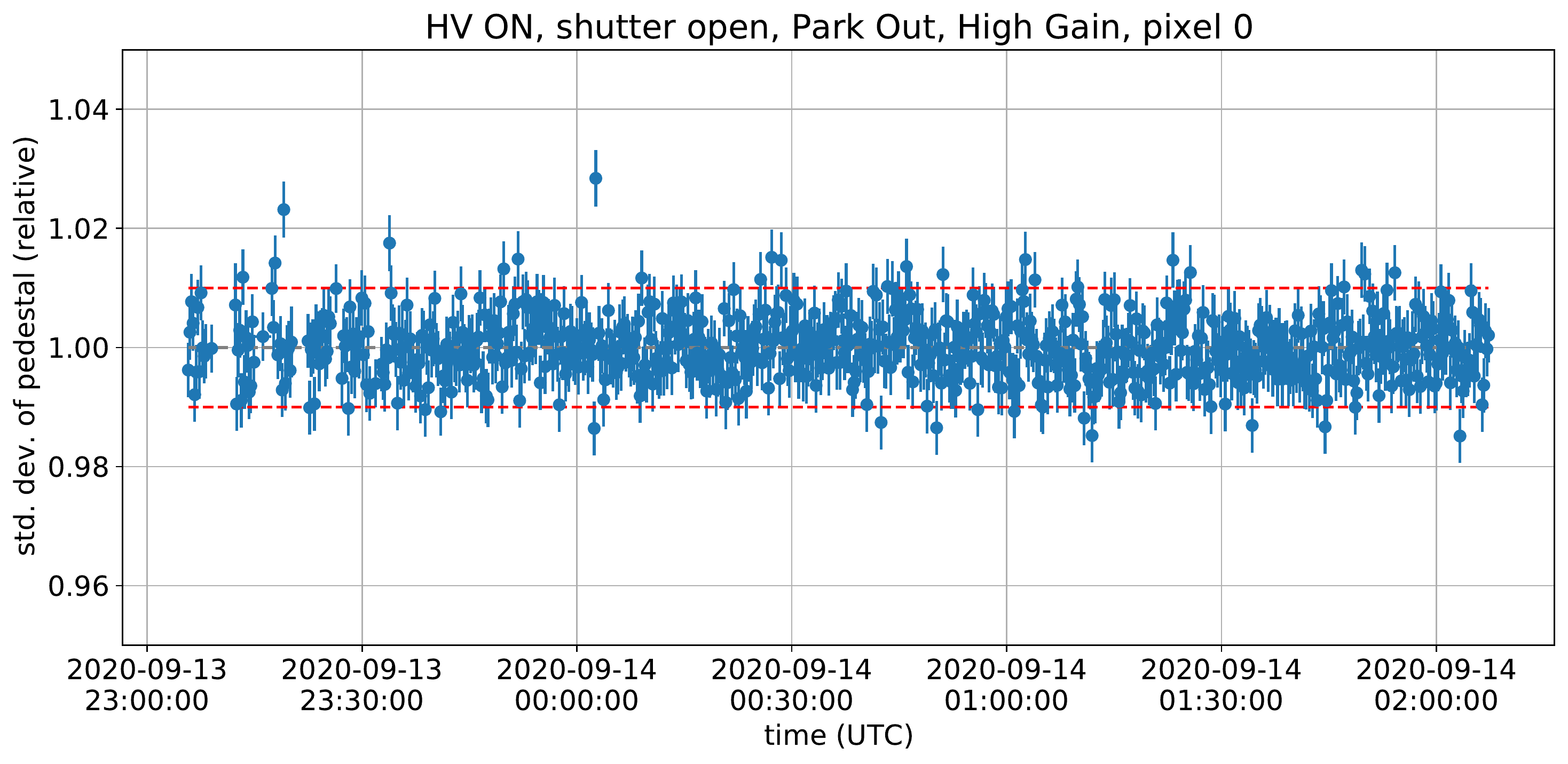}
    \caption{
    (Left) Noise level during multiple observation runs as a function of the anode current. Orange and red dotted lines show dark NSB and 10 times higher than dark NSB level, respectively. 
    (Right) Time evolution of the noise for a single pixel for three hours. Each point represents $\sim50000$ events ($\sim12$ seconds) and error bars show standard deviation of the noise for this single pixel.
    }
    \label{fig:noise}
\end{figure}

Figure~\ref{fig:noise} (Left) shows the noise level as a function of PMT anode current correlated with the night sky background (NSB) level. The noise level with the shutter closed and HVs on (blue circle) is 18.5~ADC counts corresponding to 0.21~p.e.. It is almost comparable to the electric noise without HV and the contribution of HV to the noise level is only $\sim$0.04~p.e.. During standard observations with the shutter open, the noise level gets higher with higher anode currents. A fitting function indicates that the relation between the noise level and the NSB level is explained by a square root function. The average index on the fitting function over all the pixels is $\sim$0.48. This relation is well kept even above 10 times higher NSB level than under dark condition.

Figure~\ref{fig:noise} (Right) shows the time evolution of the noise of a single pixel during a long-term (three hours) observation at park-out position. Those data points were corrected using the relation in terms of anode current seen in Figure~\ref{fig:noise} (Left). This figure indicates the noise level was stable within almost $\pm1\%$ (peak-to-peak). The average standard deviation of the noise during the long-term observation is 0.55\% in the camera even though there are some outliers possibly due to insufficient correction.

\section{PMT gain flat-fielding and calibration} \label{sec:FlatFielding}
\begin{figure}[t]
     \begin{minipage}[t]{0.49\textwidth}
      \begin{center}
       \includegraphics[width=1\textwidth]{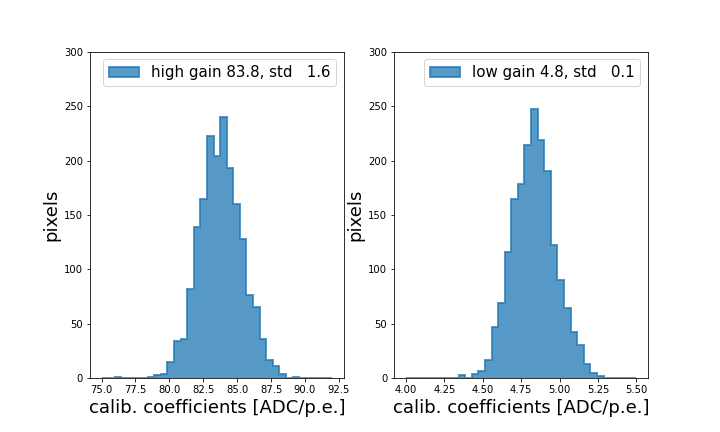}  
     \end{center}   
      \caption{Pixel calibration coefficients estimated with the F-Factor method for high gain (left) and low gain (right) channel.  }
        \label{fig:calibration_coefficients}  
    \end{minipage}
    \hfill
     \begin{minipage}[t]{0.49\textwidth}
     \begin{center}
    \includegraphics[width=0.8\textwidth]{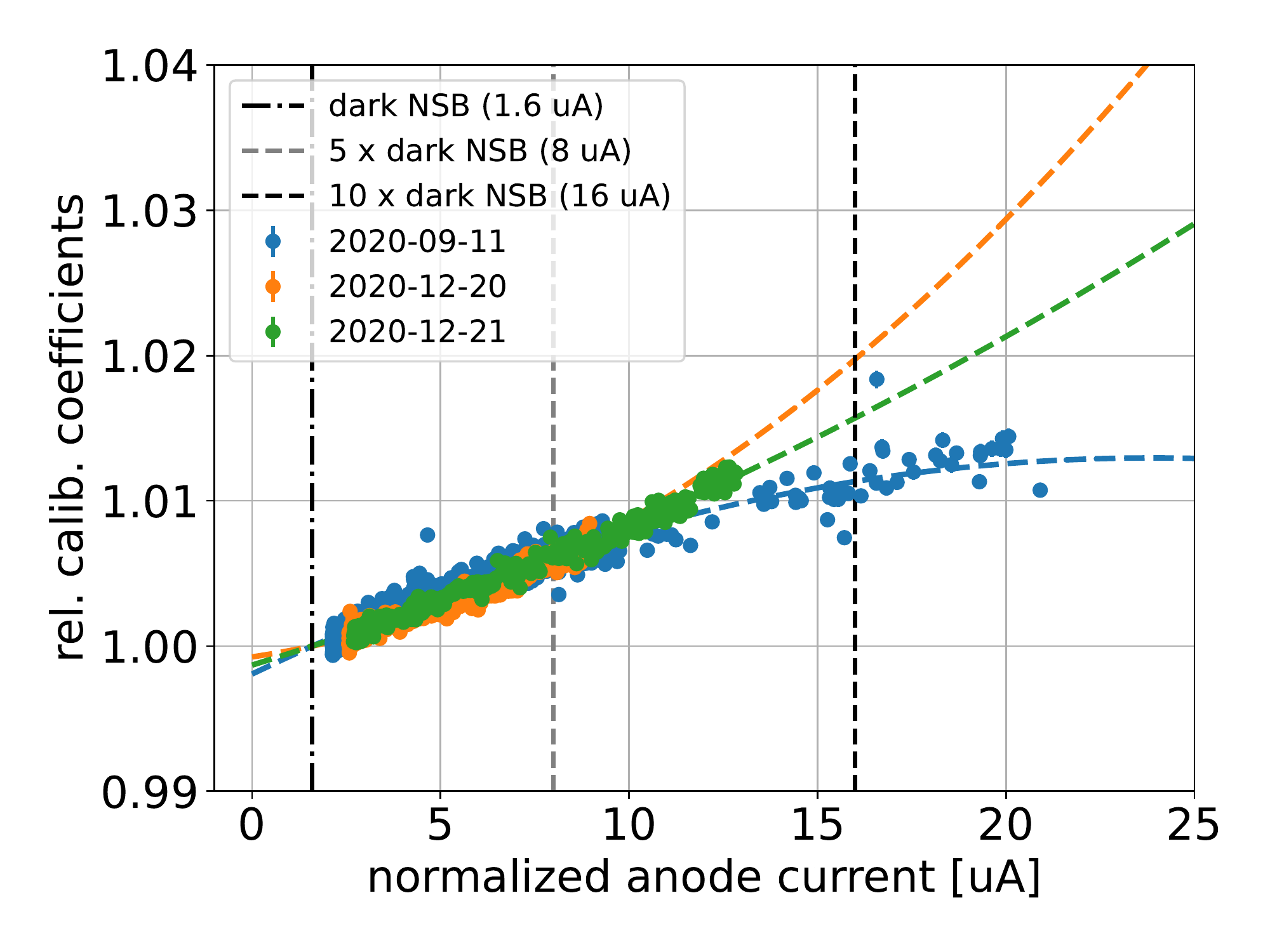}
    \end{center}
    \caption{Time evolution of relative calibration coefficients (camera average) as function of the anode current. Vertical lines indicate anode current values which correspond to different NSB levels.
    }
    \label{fig:calib_NSB}
   \end{minipage}           
\end{figure}

PMT flat-fielding and absolute gain calibration are achieved through the uniform illumination of the camera by the Calibration Box (CaliBox), placed in the center of the mirror dish. \mbox{CaliBox} consists of a hermetically closed aluminum box including a \@355 nm light laser ($\mathrm{2.4 \times 10^{12}}$ photons/pulse), a diffusive sphere and two filter wheels (6 filters each) to modulate the emitted light such as to cover the full pixel dynamic range (1-3000 p.e.). 
%
The light coming out of one of the two exits of the diffuser is sent to two sensors, a photo-diode and a Si-PM, to continuously monitor the stability of the photon flux for a high and a low intensity, respectively. The emission is measured to be stable within $\mathrm{1\%}$. The light uniformity has been verified in the laboratory to be better than 2\% scaling the camera plane at a distance of 5 m. The light intensity used for standard {\it calibration events} corresponds to $\mathrm{\sim300~ photons/pixel}$ ($\mathrm{\sim80~ p.e./pixel}$). The event rate is set to 1 kHz for specific calibration runs and 0.1 kHz for interleaved events taken during observations. \\ Pixel response equalisation is performed through high voltage flat-fielding (HV FF). The camera is illuminated with calibration events in order to estimate the relative charge of each pixel with respect to the camera median ($\mathrm{R_Q}$). Then, the relation between PMT gain G and applied voltage V
\footnote{$\mathrm{G(V)=A (V-V_1)^s+C}$ where $V_1$ = 350 V is the voltage at the first dynode and $A, s$ and $C$ are fitted parameters.} 
is used to calculate the flat-fielded voltage $\mathrm{V'}$ such that $\mathrm{G'(V') = R_Q^{-1} G(V)}$. This procedure permits to reach a charge relative dispersion (standard deviation) of less than 2\% and 3\% for the high gain and the low gain channel, respectively. \\
The estimation of absolute calibration coefficients (ADC/p.e.), per pixel and per gain channel, is achieved through the F-Factor method, which evaluates the number of photoelectrons associated to each waveform by comparing the first and the second order moments of the charge distribution. A detailed description of the procedure is given in \cite{calibICRC21}. 
Fig.~\ref{fig:calibration_coefficients} presents typical distributions of calibration coefficients for both gains. The width of the histograms reflects the residual charge dispersion after HV FF. 
The variation over time
for one pixel is close to the statistical uncertainty of the method ($\mathrm{\sim 0.5}\%$),
where samples of $10^5$ calibration events have been used for the coefficient extraction. The calibration results are influenced by the NSB up to less than 2\% at 10 times the dark NSB. Fig. \ref{fig:calib_NSB} presents the calibration coefficients as function of the anode current (dark NSB corresponds to $\mathrm{1.6~\mu A}$) for three runs taken with changing NSB, at raising or setting moon. 
%
%
%
\section{Trigger Tuning}
\label{sec:trig}
The trigger system of the LST camera consists of two trigger levels (L0 and L1), the TIB and the BPs. The L0 and L1 level are implemented in the front-end electronics.
The L0 is acting at the individual pixel level
and its circuit
allows to adjust both the times and amplitudes of the signal coming from the PMTs. Those signals are afterwards summed up for a region of 21 neighbour pixels at L1 level. 
A local camera trigger is raised if any of the summed signals is above an adjustable discriminator threshold (DT). It is propagated to the TIB through the BPs.
The TIB is managing all possible trigger inputs such as the pedestal trigger, the calibration trigger, the camera trigger of the camera itself and the ones from neighbour telescopes whenever they would exist and be connected. When the conditions are met, the trigger signal is sent back to all modules via BPs and event readout is initiated. 

\begin{figure}[htbp]
  \begin{minipage}{0.5\textwidth}
  \begin{center}
  \vspace{-0.6cm}
   \includegraphics[width=\textwidth]{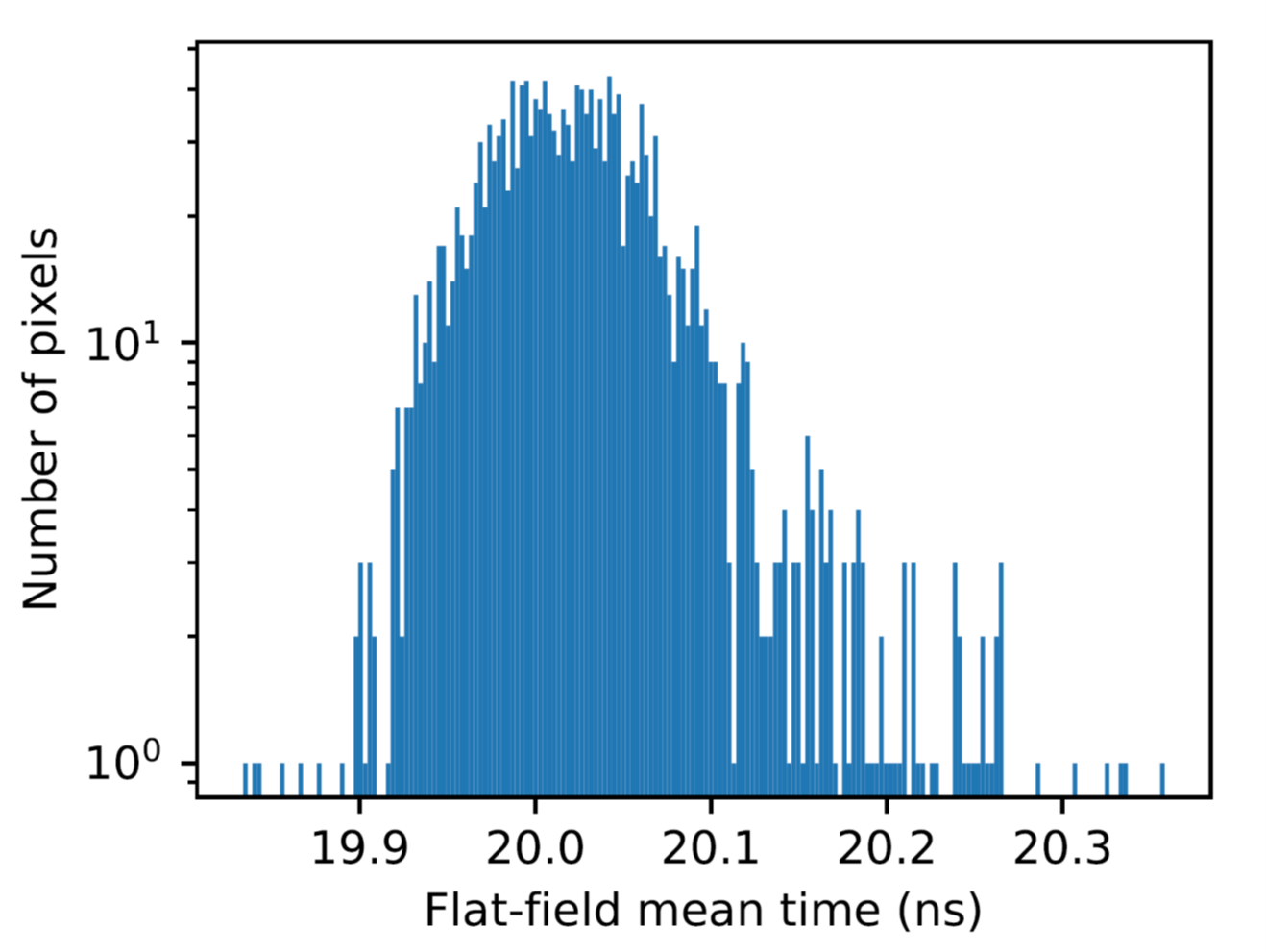}
  \end{center}
    \vspace{-0.9cm}
    \caption{Flatfield mean time distributions for all pixels.}
    \label{fig:BPTimes}
  \end{minipage}
  \hspace{0.2cm}
\begin{minipage}{0.42\textwidth}
  \begin{center}
   \includegraphics[width=\textwidth]{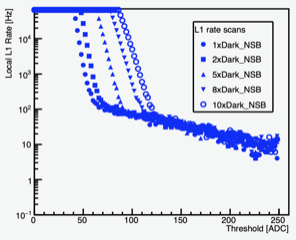}
  \end{center}
  \caption{L1 trigger rate at different NSB conditions provided by a characteristic 21 pixel trigger region as a function of the Discriminator Threshold applied.}
    \label{fig:TriggerRates}
 \end{minipage}
\end{figure}

Trigger (as well as clock) signals distribution via BPs are time-equalized in order to have a correct synchronization among all camera pixels. Since signal distribution is done in a daisy chain scheme, programmable delays are implemented inside each BP FPGA in order to achieve 1 ns equalization. These delays are calibrated by triggering the camera with the calibration box in order to have a flat-field time reference. 
Figure \ref{fig:BPTimes} shows flatfield timing performance. All pixels mean time distribution has 150~ps FWHM with an rms uncertainty of 400~ps in every pixel.
The definition of the optimal DTs is done through the rate scan procedure: the telescope is pointed to a dark patch in the sky and the trigger rate for each 21 pixel trigger region is recorded while changing the DTs.
All rate scans show a very steep wall dominated by triggers induced by the NSB and the shower region where the rate decreases much slower when increasing the DTs (figure \ref{fig:TriggerRates}). 
Then, we look for the settings that allow to get the maximum rate in all trigger regions keeping them all out of the NSB dominated region. This working point changes with the observations conditions. Stars entering the FoV affect only particular trigger regions and are taken into account by the automatic rate control (see section \ref{sec:stability}). The rate control also takes care of the change in the working point due to the zenith angle. The largest change on the working point comes from observations during moon time with an increased NSB. Figure \ref{fig:TriggerRates} shows how it changes up to a NSB 10 times larger than that in a dark condition. The working point stays inside the dynamic range for all the trigger regions even for the largest NSB in which the LST is requested to be operative (10 times the NSB of a dark condition).

The homogeneous response of the camera at trigger level is built by construction. Still, one can look at the reconstructed shower images to cross-check that homogeneity. The distribution of the position of the shower images (the center of gravity of the image) shows a standard deviation of 22\% on the number of showers in each position for the smallest recorded showers, which correspond to images with less than 50 photoelectrons, improving to 7\% for showers at analysis threshold and further for larger showers.

\section{Data Acquisition performance}

The front-end boards send data through 265 1-\Gbps\  Ethernet links, which are concentrated in six 10-\Gbps~Ethernet links by a set of six Ethernet switches. These data are received by the camera server equipped with custom tailored software for data assembly and preprocessing prior to downstream analysis and storage\footnote{The system used in LST is an evolution of the one described in: 
\cite{bib:EVB}.}. 
The protocol used is TCP/IP and the maximum transmission unit (MTU) must be lower than 2~kB.
The whole Linux network stack is replaced by the Netmap framework which intercepts data from the netwerok interface card (NIC) driver, in the kernel space, and provides it to the user space in which a deeply modified mTCP framework manages the TCP/IP algorithms.
The event fragments received from the whole camera are pre-processed. A gain selection is made on sky events and low level algorithms are applied (see below).
Events are fully assembled in processing tasks and sent to the array level processing facilities through four 10-\Gbps\ Ethernet links using 0MQ \cite{bib:0mq}.
The DAQ software running on the camera server is composed of 44 tasks running in parallel. The optimization of memory placement, of the cache usage, the lock-free synchronization approach and the avoidance of multiple copies in the network layer enable the camera server to receive and process data at full speed in real time.

Algorithms applied online to the camera data include: a pedestal subtraction,
correction of signal amplitudes depending on the time pattern of previous events and the selection of one of two linear preamplifier gains in order to optimize the dynamic range of the signals. 
In order to obtain the necessary performance on a 2-CPU-44-cores PC, the implementation has been split into 12 parallel tasks performing the time-dependent pre-processing, while 2 processing tasks alternatingly assemble these 12 fragments and apply algorithms on a complete event.



Special DAQ runs with parametrisable random and constant-frequency trigger inputs were made in order to measure the performance the complete DAQ chain from front-end electronics to downstream DAQ and data logging to be compliant with the CTA requirements for the LST, recording data up to 15~kHz and staying below 5\% of relative dead time for the complete system with event rates up to 7~kHz.
The theoretical performance of the DAQ is determined by two key values: The transmission of data from the camera through the Ethernet switch network needs 45~µs in the worst case, which limits the total input bandwidth to 19~kHz; the total output bandwidth towards the downstream DAQ is limited to 32~\Gbps\ or roughly 20~kHz event rate for data with one out of the two measured gains removed.
Thus the predominant limitation lies in the link between camera and camera server. This limit cannot be reached in practice, as the independently triggered events generate packet collisions in the network before the maximum rate is reached.
The absolute limit of DAQ bandwidth is then determined by the processing time of the software (and choice of CPUs).
This behaviour and the fulfillment of requirements are nicely illustrated by the event-rate input/output plot in Figure~\ref{fig:evbratescan}.

\vskip1pc
\vbox{\centering
  \includegraphics[width=0.7\columnwidth]{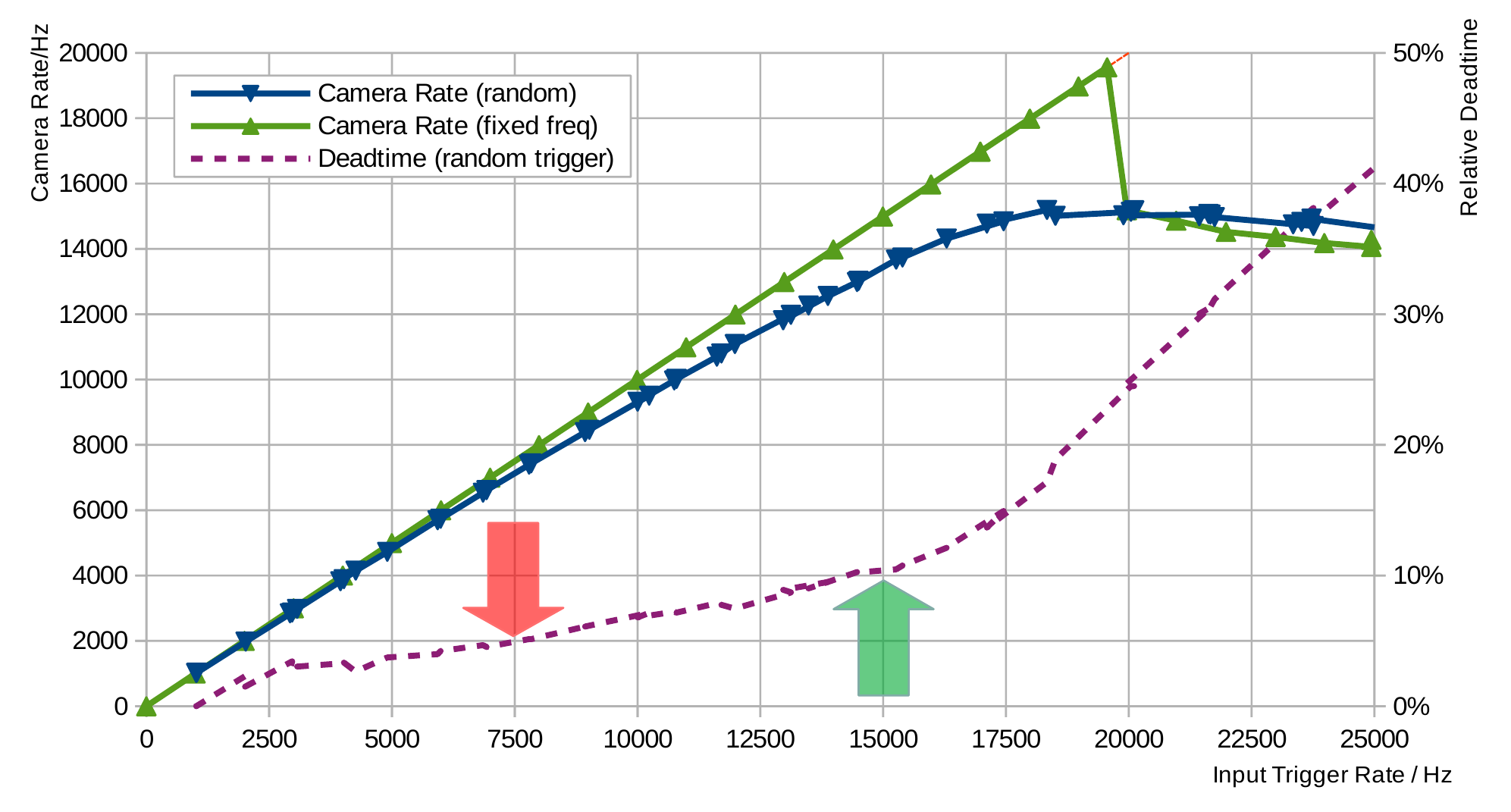}
  \captionof{figure}{\label{fig:evbratescan}
  Event output rate for randomly triggered events and triggers with constant delays (frequency) and deadtime (random trigger only) as a function of trigger input:
  The requirements on deadtime ($<5\%$ at 7~kHz, red arrow) and minimal sustainable rate (15~kHz, green arrow) are fulfilled. The deadtime lies in the linear domain for rates up to 15~kHz, significantly before saturation effects are effective.}
  }

\section{Stability of data taking} \label{sec:stability}
The low energy threshold is one of the major performance parameters of the LST. 
Among various factors, the energy threshold depends on the trigger settings of the camera modules (see Section \ref{sec:trig}).
The settings are typically chosen such that about 10--50\% of triggered events are allowed
to be due to spurious triggers caused by the fluctuations of the NSB. 
Since the NSB-caused triggers have a much steeper slope as a function of the camera threshold settings than the rate caused by air showers (see Fig \ref{fig:TriggerRates}),
choosing higher percentage of NSB triggers results in data dominated by noise. Choosing higher settings result in loss of triggers on low energy gamma-ray showers.
The LST online rate control is designed to modify the modules thresholds on the fly to ensure an operation point close to the optimal one at any reasonable NSB conditions.

The rate control has two levels. The first level acts on individual pixels, it is the so-called Level-Zero (L0) rate control.
If the pixel trigger rate is above 10 times the expected pixel rate, then this pixel is excluded from the sum-trigger.
This is mainly to exclude those pixels which are affected by bright stars. Once the rate drops below the accepted level, the pixel is enabled to participate in the sum-trigger again.
Typically 0-10 pixels are affected through the L0 rate control.

\begin{figure}[htbp]
 
 \begin{minipage}{0.47\textwidth}
  \begin{center}
   \includegraphics[width=\textwidth]{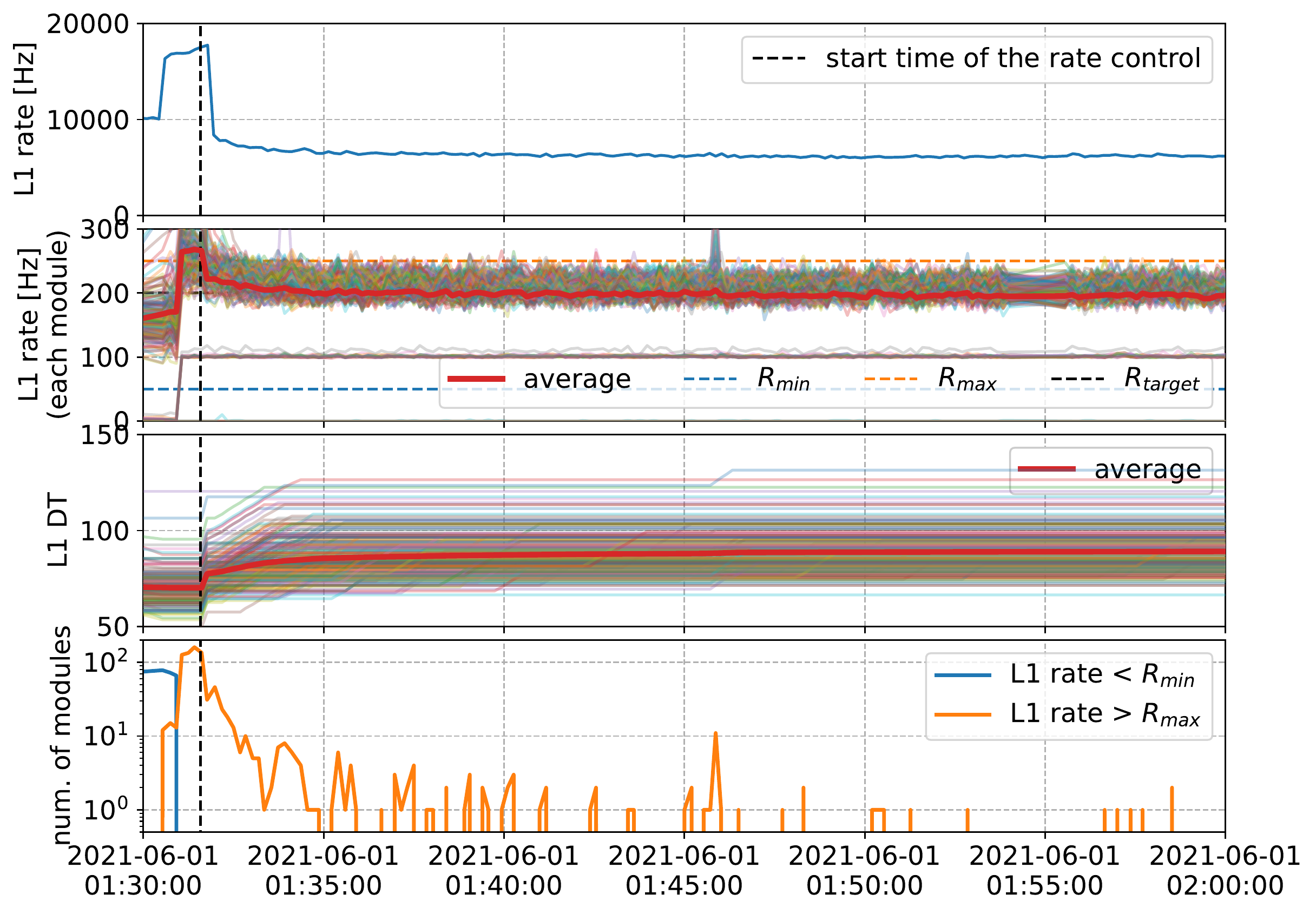}
  \end{center}
  \caption{Behaviour of the rate control. From the top, L1 total rate, L1 module rate, L1 DT, and a number of modules affected by the rate control are shown as a function of the time.}
  \label{fig:rate_control}
 \end{minipage}
\hspace{0.5cm}
 \begin{minipage}{0.45\textwidth}
  \begin{center}
   \includegraphics[width=\textwidth]{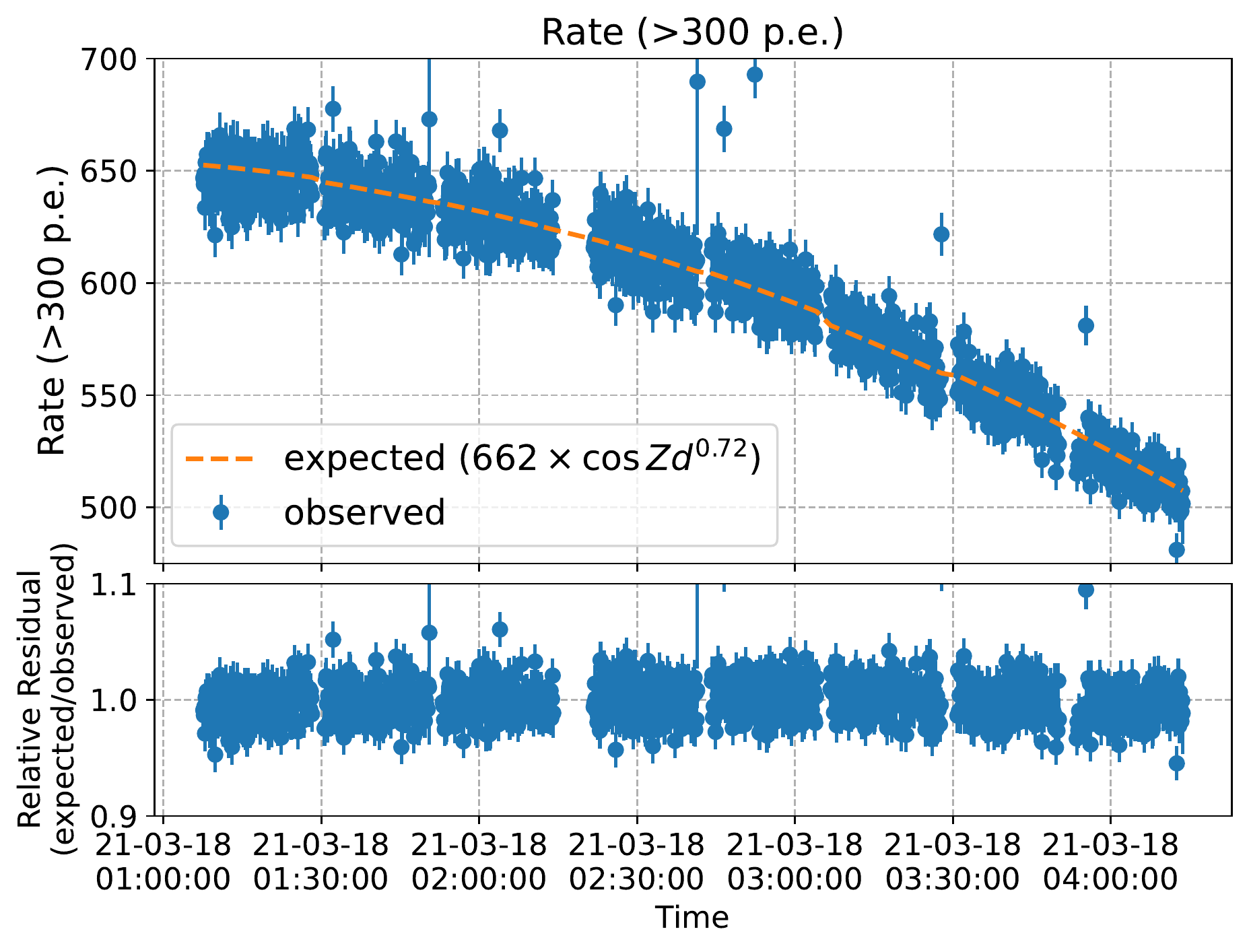}
  \end{center}
  \vspace{-0.3cm}
  \caption{Time evolution of the rates above 300~p.e. cuts after the image cleaning (Top) and the relative residual (Bottom). The orange line shows the expected rates using the zenith.}
  \label{fig:rate_stability}
 \end{minipage}
 
\end{figure}

The second level of the rate control acts on the camera module level, it is the so-called Level-One (L1) rate control. The rate R with which a given module contributes to the camera trigger
is constantly monitored and compared with two reference values: R$_{max}$ and R$_{min}$. In case R $>$  R$_{max}$, the threshold of the module is slowly increased
until the R is below a target value R$_{target}$ (with R$_{max}$ $>$ R$_{target}$   $>$ R$_{min}$). In case R is lower than the lower reference value (R $<$ R$_{min}$) the threshold is decreased until
R is above the target rate, R$_{target}$. The L1 rate control is mainly meant to reduce the effect of the faint stars in the field of view (FoV) of the camera. 
In a new FoV, when starting from default flat-fielded module thresholds,
0-40 modules are typically affected for less than 2 minutes and the corresponding module thresholds are increased. After the initial adjustment phase, typically 0-2 modules are affected due to the rotation of the stars in the FoV of the camera. Figure~\ref{fig:rate_control}
 shows an example of the behaviour of the rate control.

Figure~\ref{fig:rate_stability} shows the time evolution of the rates during the standard data taking with the rate control activated. Here, the rate is computed after the image cleaning and with a given intensity threshold (300~p.e.). It is known that the rate depends on the zenith angle and can be described as proportional to $\cos(zenith)^{\alpha}$. The power index was found to be $\alpha=0.72$ in LST-1 data after applying an image intensity cut of I$>$300\,p.e. Finally, we confirmed the rates were stable within a standard deviation of 1.5\% during the 3-hours observations taking into account the zenith dependency of the rates.

\bigskip\bigskip
{\bf Acknowledgments }

\bigskip

We gratefully acknowledge financial support from the agencies and organizations listed here:  www.cta-observatory.org/consortium\_acknowledgments”

\clearpage
\section*{Full Authors List: \Coll\ Collaboration}
\scriptsize
\noindent
H. Abe$^{1}$,
A. Aguasca$^{2}$,
I. Agudo$^{3}$,
L. A. Antonelli$^{4}$,
C. Aramo$^{5}$,
T.  Armstrong$^{6}$,
M.  Artero$^{7}$,
K. Asano$^{1}$,
H. Ashkar$^{8}$,
P. Aubert$^{9}$
% ...
\\

\noindent
$^{1}$Institute for Cosmic Ray Research, University of Tokyo.
$^{2}$Departament de Física Quàntica i Astrofísica, Institut de Ciències del Cosmos, Universitat de Barcelona, IEEC-UB.
$^{3}$Instituto de Astrofísica de Andalucía-CSIC.
$^{4}$INAF - Osservatorio Astronomico di Roma.
$^{5}$INFN Sezione di Napoli.
$^{6}$Aix Marseille Univ, CNRS/IN2P3, CPPM.
$^{7}$Institut de Fisica d'Altes Energies (IFAE), The Barcelona Institute of Science and Technology.
$^{8}$IRFU, CEA, Université Paris-Saclay.
$^{9}$LAPP, Univ. Grenoble Alpes, Univ. Savoie Mont Blanc, CNRS-IN2P3, Annecy.
% ...

%
%

\end{document}